\documentclass[11pt]{article} 
\usepackage{hyperref} 
\pdfoutput=1 
 
\begin{document} 
 
\title{Falling Jets of Particles in Viscous Fluids} 
 
\author{Florent Pignatel$^{1}$, Maxime Nicolas$^{1}$, \'Elisabeth Guazzelli$^{1}$,\\ and David Saintillan$^{2}$ \vspace{6pt} \\ 
$^{1}$ IUSTI CNRS UMR 6595, Polytech' Marseille, \\ Aix-Marseille Universit\'e (U1), France \vspace{4pt} \\
$^{2}$ Department of Mechanical Science and Engineering, \\ University of Illinois at Urbana-Champaign, USA} 
 
\maketitle 
 
 
\begin{abstract} 
This fluid dynamics video presents experiments and simulations of gravity-driven particulate jets in viscous fluids at low Reynolds number. An initially straight jet is shown to develop varicose modulations of its diameter as it sediments under the action of gravity. While this instability is qualitatively reminiscent of the classical Rayleigh-Plateau instability for immiscible fluids, its mechanism has yet to be understood as neither inertia nor surface tension play a role in the case of a dilute suspension at $Re=0$. \end{abstract} 
 
 
\section{Discussion} 

Experiments on gravity-driven suspension jets were performed by letting a dilute suspension of spheres ($\phi\sim 1\%$ to $10\%$) settle by gravity out of a tube into a tank containing the same viscous liquid as that used to prepare the suspension. The density of the particles and the liquid viscosity were chosen so as to make the Reynolds number low ($Re\sim 10^{-3}$). Simulations of such jets were also performed in the Stokes flow regime using a point-particle approximation.

This fluid dynamics \href{http://ecommons.library.cornell.edu/bitstream/1813/11475/2/FilmAPS_19septMPEG2.mpg}{video} shows a typical experiment together with the corresponding simulation. Good qualitative agreement is observed between experiment and simulation, and both exhibit an instability by which an initially straight jet develops varicose modulations of its diameter. These modulations are reminiscent of the classical Rayleigh-Plateau instability of immiscible jets; in the present case, however, neither inertia nor surface tension play a role, and the precise mechanism of the instability in low-$Re$ suspensions remains unknown.
 \end{document}